\documentclass[twocolumn,amsmath,letterpaper,amssymb,prl,floatfix]{revtex4-1}

\usepackage{graphicx}
\usepackage[font=small, justification=centerlast]{caption}
\usepackage{color}

\newcommand{\mb}[1]{\mathbf{#1}}
\newcommand{\hmb}[1]{\hat{\mathbf{#1}}}

\begin{document}
\bibliographystyle{nature1}

\title{Plasticity of hydrogen bond networks regulates mechanochemistry of cell adhesion complexes}

\author{Shaon Chakrabarti, Michael Hinczewski
  and D. Thirumalai} \affiliation{Biophysics Program, Institute For
  Physical Science and Technology, University of Maryland, College
  Park, MD 20742}

\begin{abstract}
 Mechanical forces acting on cell adhesion receptor proteins regulate a range of cellular functions by formation and rupture of non-covalent interactions with ligands. Typically, force decreases the lifetimes of intact complexes (slip-bonds), making the discovery that these lifetimes can also be prolonged (``catch-bonds"), a surprise. We created a microscopic analytic theory by incorporating the structures of selectin and integrin receptors into a conceptual framework based on the theory of stochastic equations, which quantitatively explains a wide range of experimental data (including catch-bonds at low forces and slip-bonds at high forces).  Catch-bonds arise due to force-induced remodeling of hydrogen bond networks, a finding that also accounts for unbinding in structurally unrelated integrin-fibronectin and actomyosin complexes. For the selectin family,  remodeling of hydrogen bond networks drives an allosteric transition resulting in the formation of maximum number of hydrogen bonds determined only by the structure of the receptor and is independent of the ligand. A similar transition allows us to predict the increase in number of hydrogen bonds in a particular allosteric state of $\alpha_5 \beta_1$ integrin--fibronectin complex, a conformation which is yet to be crystallized. We also make a testable prediction that a single point mutation (Tyr51Phe) in the ligand associated with selectin should dramatically alter the nature of the catch-bond compared to the wild type. Our work suggests that nature utilizes a ductile network of hydrogen bonds to engineer function over a broad range of forces.
  \end{abstract}

\maketitle
\def\s{\rule{0in}{0.28in}}

\section*{Significance Statement}

Selectins and integrins are receptor proteins on cell surfaces
responsible for adhesion to a variety of extracellular biomolecules,
a critical component of physiological processes like white blood cell
localization at sites of inflammation.  The bonds which the receptors
form with their targets are regulated by mechanical forces (for example
due to blood flow). The bond lifetimes before rupture exhibit a counterintuitive
increase with force, before decreasing as expected. Based on crystal
structures of selectin and integrin, we created a general analytic
theory which for the first time relates microscopic structural
rearrangements at the receptor-ligand interface to macroscopic bond
lifetimes.  We quantitatively explain experimental data from diverse
systems spanning four decades of lifetime scales, and also predict the
outcome of mutations in specific residues.

\section*{Introduction}

Cells communicate with each other and their surroundings in order to
maintain tissue architecture, allow cellular movement, transduce
signals and heal wounds~\cite{berrier_cell-matrix_2007}. Important
components in many of these processes are cell adhesion
molecules---proteins on cell surfaces that recognize and bind to
ligands on other cells or the extracellular matrix
\cite{berrier_cell-matrix_2007, gumbiner_cell_1996}. For example,
adhesion of leukocytes to the endothelial cells of the blood vessel is
a vital step in rolling and capturing of blood cells in wound-healing,
and is mediated by the selectin class of receptor
proteins~\cite{ley_getting_2007}. The functional responses of cell
adhesion molecules are often mechanically transduced by shear stresses
and forces arising from focal adhesions to the cytoskeleton or simply
the flow of blood in the vasculature. Under stress, molecules undergo
conformational changes, triggering biophysical, biochemical, and gene
regulatory responses that have been the subject of intense
research~\cite{davies_flow-mediated_1995,
  traub_laminar_1998}. Lifetimes of adhesion complexes are typically
expected to decrease as forces
increase~\cite{bell_models_1978}. However, the response of certain
complexes to mechanical force exhibits a surprisingly counterintuitive
phenomenon. Lifetimes increase over a range of low force values,
corresponding to ``catch-bond''
behavior~\cite{dembo_reaction-limited_1988}.  At high forces, the
lifetimes revert to the conventional decreasing behavior,
characteristic of a ``slip-bond''~\cite{bell_models_1978}.  In
retrospect, the plausible existence of catch-bonds was already evident
in early experiments by Greig and Brooks, who discovered that
agglutination of human red blood cells using the lectin concanavalin
A, increased under shear~\cite{greig_shear-induced_1979}. Although not
interpreted in terms of catch-bonds, their data showed lower rates of
unbinding with increasing force on the complex. Given the importance
of mechanotransduction in cellular adhesions, a quantitative and
structural understanding of this surprising phenomenon is imperative.
 
Direct evidence for  catch-bonds in  a wide variety of cell
adhesion complexes have come from flow and AFM experiments in the last
decade \cite{thomas_bacterial_2002, marshall_direct_2003,
  kong_demonstration_2009}, along with examples from other
load-bearing cellular complexes like actomyosin
bonds~\cite{guo_mechanics_2006} and microtubule-kinetochore
attachments~\cite{akiyoshi_tension_2010}.  The catch-bond lifetime
exhibits non-monotonic biphasic behavior---increasing up to a certain
critical force and decreasing at larger forces. The structural
mechanisms leading to catch-bond behavior have largely been elusive,
though experiments have provided key insights for
selectins~\cite{somers_insights_2000,phan2006} and
integrins~\cite{luo_structural_2007,zhu_complete_2013}.  In these
systems, the rupture rate of the ligand from the receptor depends on
an angle between two domains in the receptor molecule
(Fig.~\ref{f1}). Conformations with smaller angles detach more slowly
than those with large angles. In the case of integrins, multiple
conformations at varying angles have been
crystallized~\cite{zhu_complete_2013}, while for selectins, only two
(see Fig.~\ref{f1}) have been found so far in the crystal
structures~\cite{somers_insights_2000}. In the absence of an external
force, the molecule fluctuates between conformations corresponding to
a variety of angles, including the larger angles from which the ligand
can rapidly detach.  With the application of force, the two domains
increasingly align along the force direction, restricting the system
to small angles and longer lifetimes, until large forces again reduce
the barrier to rupture.

Previously, theories based on kinetic models with the assumption of a
phenomenological Bell-like coupling of rates to
force~\cite{evans_mechanical_2004, barsegov_dynamics_2005,
  lou_flow-enhanced_2006,Pereverzev2009} have been used to explain
catch-bond behavior. However, the parameters extracted from these kinetic
models cannot be easily related to microscopic physical processes in
specific catch-bond systems.  More importantly, such models merely
rationalize the experimental data, and do not have predictive
power. The large scale of catch-bond lifetimes, $\sim 10-10^4$ ms,
makes it impossible to directly observe unbinding in a realistic
all-atom simulation, much less the macroscopic consequences of
mutations. 
 
Here, we solve the difficulties alluded to above by creating a new
theoretical approach.  By building on the insights from the structures
of cell adhesion complexes, we introduce a microscopic theoretical
model that captures the essential physics of the angle-dependent
detachment, and its implications for catch-bond behavior.  Taking cue
from the crystal structures of selectin and integrin, we construct a
coarse-grained energy function for receptor-ligand interactions.  The
model yields an analytic expression for the bond lifetime as a
function of force, which gives excellent fits to a broad range of
experimental data on a number of systems.  The extracted parameters
have clear structural interpretations, and their values provide
predictions for energetic and structural features like strength of
hydrogen bonding networks at the receptor-ligand interface.  Where
estimates of these properties can be directly obtained from crystal
structures, our predictions are in remarkable agreement. The energy
scales identified through the model are specific enough to allow
predictions for structures not yet crystallized, and suggest novel
mutation experiments that would modify catch-bond behavior in
quantifiable ways.  For the selectins, we predict how a specific
mutation in the PSGL-1 ligand will alter its unbinding from P-selectin
under force, and provide new interpretation of data from L-selectin
mutants~\cite{lou_flow-enhanced_2006}. Interestingly, the experimental
fits suggest that both P- and L-selectin have a characteristic,
ligand-independent energy scale, determined by the chemistry of their
binding interfaces. For integrins, we predict the strength of extra
interactions that should be observed in a crystal structure of the
$\alpha_5 \beta_1$--fibronectin complex in an open state. The
generality of the theory is further established by obtaining
quantitative agreement for the catch-bond behavior in actomyosin
complex. Our theory provides the first structural link between the
catch to slip bond transition in cell adhesion complexes, covering
a broad range of forces and lifetimes.

\section*{Theory}

\begin{figure}[t]
 \includegraphics[width=\columnwidth]{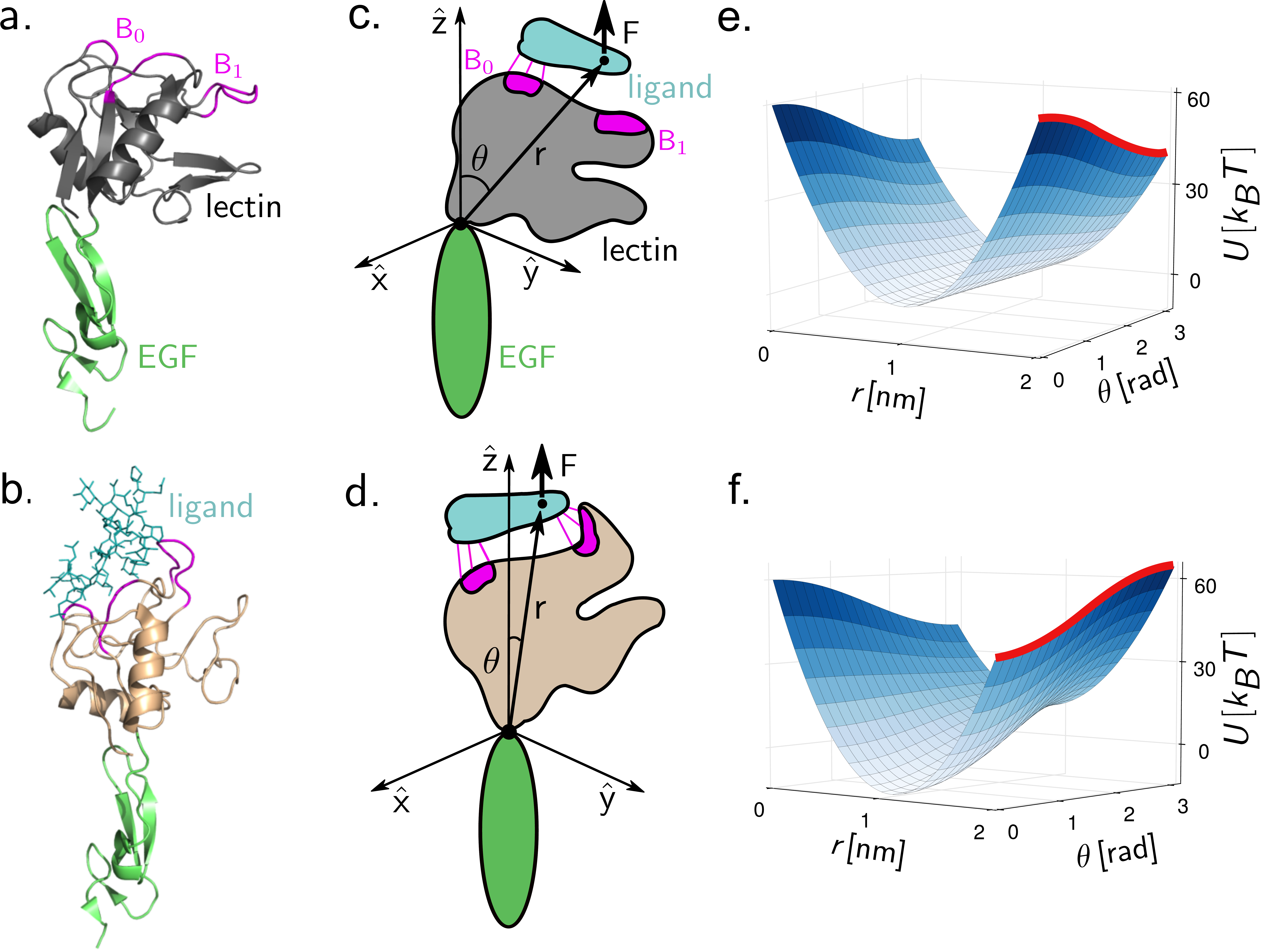}
  \caption{Abstraction of the model based on structure. a. Crystal structure of
    P-selectin~\cite{somers_insights_2000} in the bent conformation
    (PDB: 1G1Q); b. the extended conformation (PDB: 1G1S).  The lectin
    (gray/beige) and EGF (green) domains are labeled, along with two
    regions of the ligand binding interface (B$_0$ and B$_1$, purple).
    The ligand (an N-terminal fragment of the glycoprotein PSGL-1) is
    only co-crystallized in the extended state.  c-d. Schematic
    conformations of our model, corresponding to panels a and b.  e-f.
    Plots of the potential $U(r,\theta)$ at $F=0$ and $F=50$pN
    respectively, with $k_0=80 \, k_BT/\text{nm}^2$, $k_1=20 \,
    k_BT/\text{nm}^2$, $r_0=1.0 \, \text{nm}$, and $b=2$ nm.  The
    energy $U(b,\theta)$ at the transition state is highlighted
    in red.}\label{f1}
\end{figure}

\begin{figure}[t]
 \includegraphics[width=\columnwidth]{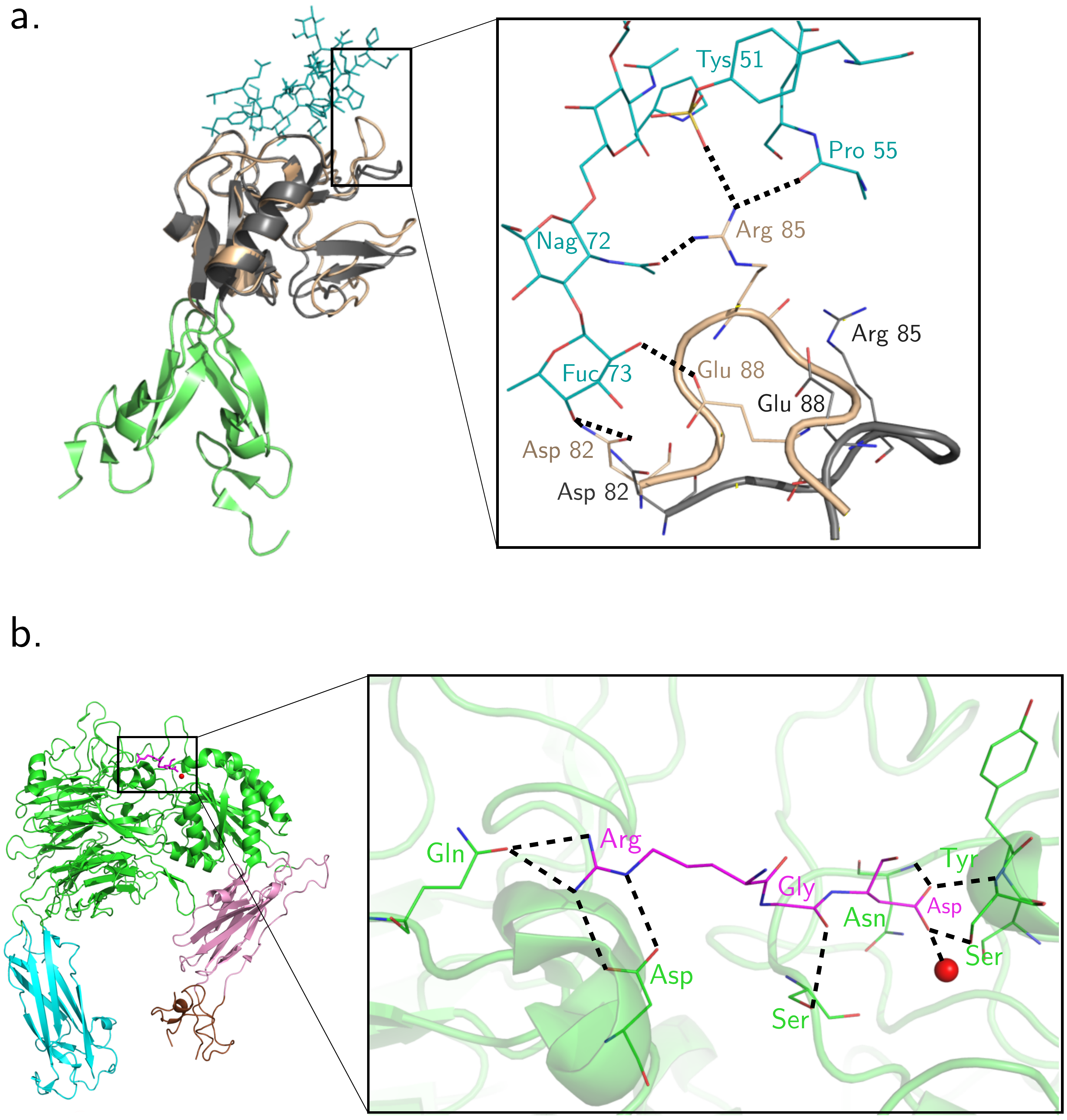}
  \caption{Receptor--ligand hydrogen bond networks in P-selectin and $\alpha_5 \beta_1$ integrin. a. The crystal structures from Fig.~\ref{f1}a,b
    superimposed with aligned lectin domains.  The inset shows the
    remodeling of the $B_1$ region of the ligand-binding interface
    (the Asp82--Glu88 loop).  In the extended state (beige) this loop
    forms a network of hydrogen bonds (dashed lines) with the ligand
    (to be compared with $E_1$ of our model).  In the bent state
    (gray) the loop rotates sufficiently far that it is unlikely to
    participate in
    binding~\cite{somers_insights_2000,springer_structural_2009}. b. Hydrogen
    bond network between ligand RGD and $\alpha_5 \beta_1$ integrin in
    the closed headpiece conformation (PDB: 3VI4). The integrin
    domains are colored as follows: $\beta$-propeller and $\beta$A in
    green (cartoon and line representation), thigh in cyan, hybrid in
    pink and PSI in brown. The ligand is colored magenta (stick or
    line) and the MIDAS magnesium ion is red (sphere).  This network
    should be compared to $E_0$, predicted from our model.}\label{f2}
\end{figure}

\noindent\textbf{Structural basis of catch-bond model:} We now define our model
using the structures of P-selectin, which has been crystallized in two
conformations: a ``bent'' [Fig.~\ref{f1}a] and ``extended'' state
[Fig.~\ref{f1}b]~\cite{somers_insights_2000}. The two states differ by
the angle which the EGF domain (green in Figs.~\ref{f1}a
and~\ref{f1}b) assumes with respect to the lectin domain (gray/beige).
Although ligands can bind to lectin in both conformations,
co-crystallization with the ligand (the truncated N-terminal portion
of the glycoprotein PSGL-1) was achieved only for the extended state.
In this latter case, there are two major regions of the lectin domain
(B$_0$ and B$_1$, colored purple in Figs. 1a - 1d) that form substantial hydrogen bond
networks with the ligand, thereby stabilizing the complex.  Based on
alignments of the lectin domain in the bent and extended states
[Fig.~\ref{f2}a], it is believed that the binding interface is
remodeled in the bent
conformation~\cite{somers_insights_2000,springer_structural_2009}.
The region B$_1$ (a loop between Asp82 and Glu88, shown in the inset
of Fig.~\ref{f2}a) rotates so that it can no longer engage the ligand.
This angle-dependent rearrangement results in weaker ligand
attachment, and hence explains the shorter bond lifetimes in the bent
vs. the extended conformation.

Our minimal model, which captures the structure based angle-dependent
dissociation, describes the ligand-receptor interaction through an
effective spring with bond vector $\mb{r} \equiv \left( r,\theta,\phi
\right)$ between a pivot point within the receptor and a point in the
ligand [Fig.~\ref{f1}c-d].  The pivot point is fixed at the origin,
and the ligand is under an external force $F \hmb{z}$ along the
$z$-axis.  The energy associated with the spring is given by the potential,
\begin{equation}\label{e1}
U(r,\theta)=\frac{1}{2}\left( k_{0}+k_{1}(1+\cos\theta) \right)  (r-r_0)^{2} - F r\cos\theta,
\end{equation}
with $k_0,k_1>0$.  The first term is an elastic energy, where $r_0$ is
the natural length of the bond magnitude $r$, and
$k_{0}+k_{1}(1+\cos\theta) \equiv k(\theta)$ is an angle-dependent
spring constant.  The second term is the contribution due to the
mechanical force $F$.  The bond ruptures if $r \ge b$, where $b \equiv
r_0 +d$ and $d$ is the transition state distance.  The structural
inspiration of our model naturally leads to a multidimensional
potential energy, dependent on both $r$ and $\theta$, which is a key
requirement for a physically sensible description of catch-bond
behavior.  Any one-dimensional potential with rupture defined by a
single cutoff in some reaction coordinate will always exhibit
monotonic lifetime behavior as a function of force.  We assume that
the time evolution of the vector $\mb{r}$ follows a Fokker-Planck
equation, describing diffusion on the potential surface $U(r,\theta)$
with a diffusion constant $D$.  We define the lifetime of the bond
$\tau(F)$ at a given force $F$ as the mean first passage time from
$\mb{r}_\text{min}(F)$, the position of the minimum in $U$, to any
$\mb{r}$ with $r=b$.  Implicit in this diffusive picture is the
assumption that the angle $\theta$ can change continuously.  This is a
reasonable approximation even if the receptor-ligand complex
fluctuates between several discrete angular
states~\cite{zhu_complete_2013}, assuming the energy barriers between
the states are such that the interconversion between states happens on
much faster timescales than $\tau(F)$.  The presence of the barriers
would in this case be incorporated through a renormalization of the
effective diffusion constant $D$~\cite{Zwanzig1988}.

We show in Fig.~\ref{f1}e a representative zero-force potential energy
surface $U(r,\theta)$, with the energy at rupture $U(b,\theta)$,
highlighted in red. The form of $k(\theta)$ makes it energetically
favorable for bond rupture at $\theta =\pi$ (the bent state), with a
cost $E_0 = k_0 d^2/2$ to dislodge the ligand to the failure point.  In
the opposite limit of $\theta =0$ (the extended state), energy for
rupture is highest, with a cost $E_0 + E_1$, where $E_1 = k_1 d^2$.
The values of $E_0$ and $E_1$ correspond to the stabilization energies
associated with the ligand in the two allosteric (bent and extended)
states. However, as $F$ is increased, the bond aligns along $\hmb{z}$,
and the minimum in $U(r,\theta)$ shifts toward $\theta =0$
[Fig.~\ref{f1}f], biasing the system toward the extended state.  Thus,
we expect the lifetime $\tau(F)$ to initially increase with $F$ and
eventually decrease at forces sufficiently large to reduce the rupture
barrier.

Though the schematic diagram of the model in Fig.~\ref{f1}c-d draws
the vector $\mb{r}$ between a pivot at the EGF-lectin interface to the
tip of the ligand, one should note that the actual ligand-lectin
complex does not behave like a perfectly rigid object rotating about a
hinge, nor does it cover the entire angular range between $\theta=0$
and $\pi$. Since proteins are deformable, the pivot location and the
length $r_0$ will depend on the compliance of the specific domains
involved in reorientation. Hence, we expect $r_0$ to be of the order
of, or less than the size of the localized domains that rotate to
become restructured under force. It therefore follows that the
structures of the complex in different allosteric states provide
valuable insights into their response to force.  The length scale $d$
reflects the brittleness of the bonding interactions~\cite{Hyeon2007},
with larger $d$ indicating a malleable bond interface which can be
deformed over longer distances before the complex falls apart.  The
two energy scales $E_0$ and $E_1$ also have physical interpretations,
with $E_0$ being roughly the total strength of noncovalent
interactions between the region $B_0$ and the ligand, whereas $E_1$ is
the additional contribution from the region $B_1$ in the extended
conformation.

\section*{Results and Discussion}

\noindent\textbf{Mean bond lifetime:} By assuming that
$\tau(F)$ is much longer than the local equilibration time around
$\mb{r}_\text{min}(F)$, the lifetime of the complex is approximately given by,
\begin{equation}\label{e2} \tau (F)\approx \frac{ \sqrt{\pi}
    \, r_0 (E_1 - 2
    F (d + r_0)) e^{\beta(E_0+d F)} (e^{2 \beta F r_0}-1)}{ 4 D (\beta E_0)^{3/2} F \left(1+ r_0/d \right) ^2 (1 - e^{\beta (2F (d + r_0)-E_1)})},
\end{equation}
where $\beta = 1/k_B T$, $k_B$ is the Boltzmann constant, and $T$ is
the temperature.  The full details of the derivation are in the
SI. The central result of this work in Eq.~\eqref{e2}
provides an analytic expression for the mean first passage time in
terms of the microscopic energy and length scales covering both the
catch and slip-bond regimes. To set a reasonable scale for $D$, we
make it equal to the diffusivity of a sphere of radius $r_0$, $D=k_B
T/6 \pi \eta r_0$, where $\eta$ is the viscosity of water.  (A
prefactor in $D$ due to molecular shape can be absorbed as a small
logarithmic correction to the energy scale $E_0$.)  With this
assumption, Eq.~\eqref{e2} becomes an equation with four parameters:
$E_0$, $E_1$, $d$ and $r_0$.  We validated the approximation
underlying Eq.~\eqref{e2} by comparison to Brownian dynamics
simulations of diffusion on $U$ (details in the SI), which showed
excellent agreement with our analytical theory (see Fig. S2 in the
SI).

For $\beta F d \gg 1$, $\tau(F)$ decays exponentially in a manner similar to
the standard Bell model for systems exhibiting slip-bonds, $\tau(F) \sim
\exp(-\beta dF)$. The decay rate is controlled by the transition state
distance $d$.  The characteristic catch-bond behavior occurs at
smaller $F$, where we see a biphasic $\tau(F)$ peaked at $F =
F_\text{p}$,
\begin{equation}\label{e3}
  F_{\text{p}} \approx \frac{A E_1}{2(r_0+d)},
\end{equation}
with a prefactor $A \sim {\cal O}(1)$.  The ratio of the peak height
$\tau(F_\text{p})$ to the lifetime $\tau(0)$ at zero force, which is a
measure of the strength of the catch-bond, scales like
\begin{equation}\label{e4}
  \frac{\tau(F_\text{p})}{\tau(0)} \approx \frac{4 A^\prime (d+r_0)}{r_0E_1^2} \sinh\left(\frac{E_1}{2}\right) \sinh \left(
    \frac{r_0 E_1}{2(d+r_0)} \right),
\end{equation}
with a prefactor $A^\prime \sim {\cal O}(1)$.  From Eqs.~\eqref{e3}
and \eqref{e4} we see that $E_1 \to 0$ leads to $F_\text{p}\to 0$ and
$\tau(F_\text{p})/\tau(0) \to 1$.  In this limit, the model predicts only
slip-bond behavior, where the lifetime decreases monotonically with
force.  Thus, our model interpolates between catch-bond and slip-bond
regimes by varying the energy scale $E_1$.

\begin{figure}[t]
 \includegraphics[width=\columnwidth]{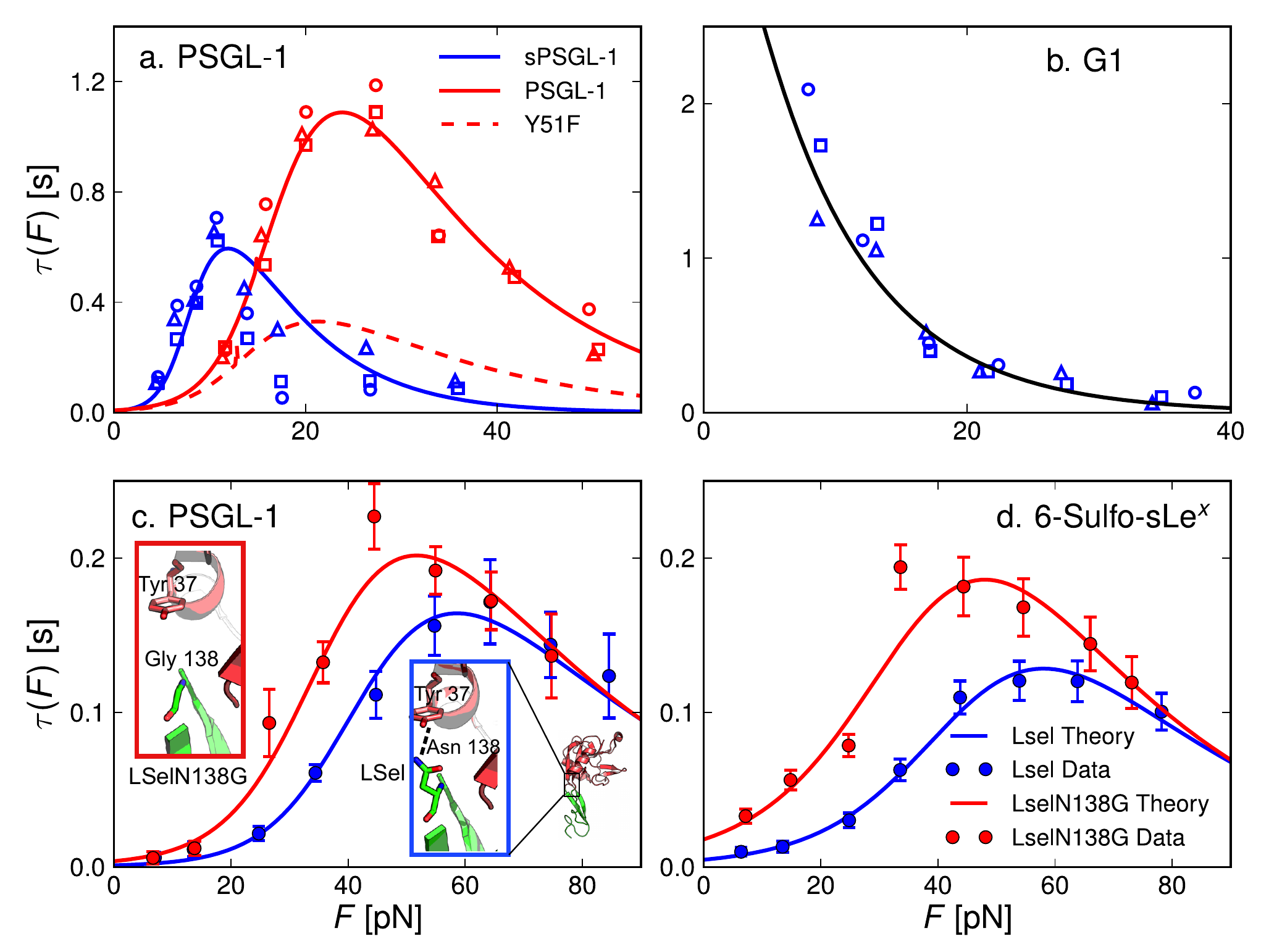}
\caption{Experimental best-fit results for bond lifetime $\tau(F)$
  versus force $F$ for selectins.  The top and bottom rows correspond
  to the receptors P-selectin (Psel) and L-selectin (Lsel)
  respectively. The ligands are indicated above the figures.  The
  symbols are experimental results and the lines are analytical curves
  from Eq.~\eqref{e2}, with parameters given in Table~I.  The sources
  for the data are: a-b) \cite{marshall_direct_2003}; c-d)
  \cite{lou_flow-enhanced_2006}.  For panels a-b, the symbol shapes
  denote three alternate ways of estimating experimental $\tau(F)$.
  Squares: average of the lifetimes; triangles: standard deviation of
  the lifetimes; circles: -1/slope in the logarithmic plot of the
  number of events with lifetime $t$ or more versus $t$.  Up to
  sampling errors, these estimates are equivalent for systems with
  exponentially distributed bond lifetimes.}\label{f3}
\end{figure}

\begin{figure}[t]
\includegraphics[width=\columnwidth]{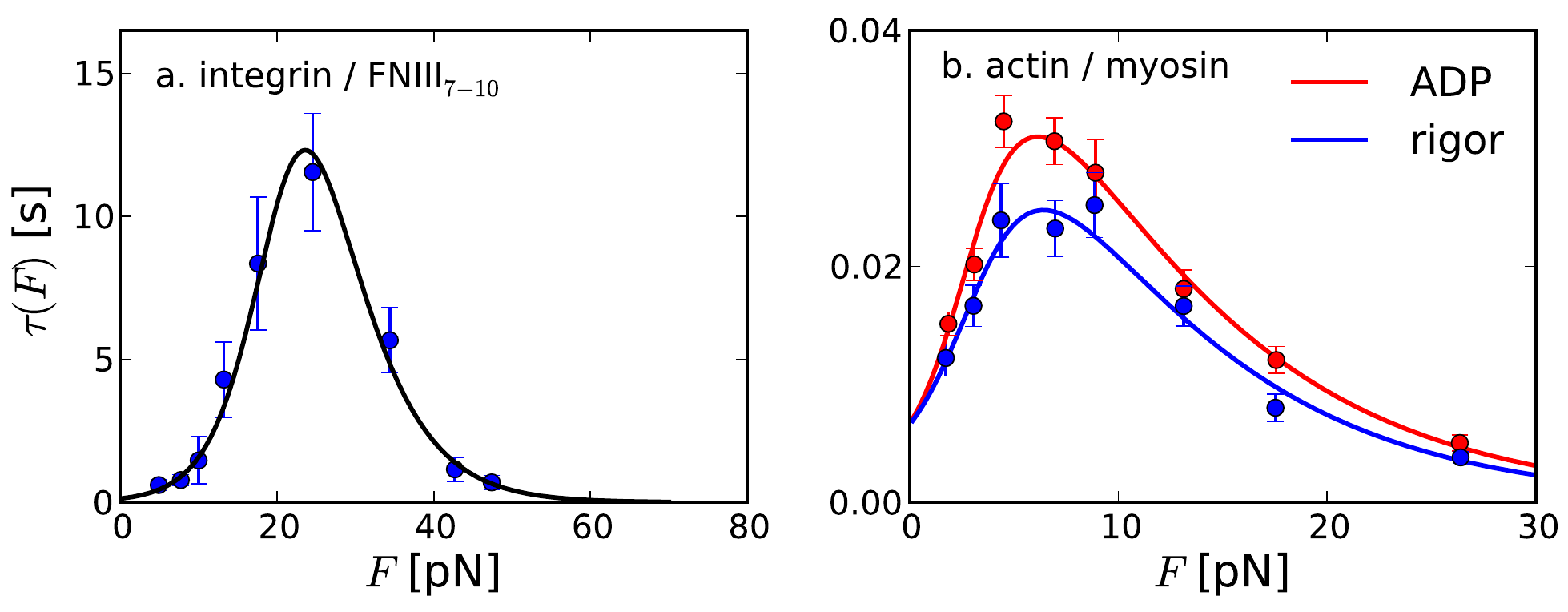}
\caption{Experimental best-fit results for bond lifetime $\tau(F)$ for
  non-selectin complexes. The receptor ligand systems are indicated on
  top of the figures. Sources of the data are
  a.~\cite{kong_demonstration_2009} and b.~\cite{guo_mechanics_2006}. }\label{f4}
\end{figure}

\noindent\textbf{Analysis of experimental data:} We first establish the
efficacy of the theory by analyzing experimental data for $\tau(F)$
for a variety of complexes. The fits in Fig.~\ref{f3} (selectins) and
Fig.~\ref{f4} (non-selectin complexes) show that there is excellent
agreement between the analytical theory and measurements, which is
remarkable since our microscopic model shows that only a small number
of fitting parameters suffice to fit nine complexes with vastly
differing architectures. These experiments involve applying force to
molecular complexes either through AFM or optical traps, with the
force initially ramped from zero to a given value $F$.  Bonds which
survive the ramp are then held at constant $F$ until rupture.  If the
initial ramp is sufficiently slow such that the system always remains
quasi-adiabatically in equilibrium at the instantaneous applied
force~\cite{raible2004}, the subsequent duration of the bond while at
constant $F$, averaged over many trials, provides an accurate estimate
of $\tau(F)$.  (Extremely high ramp speeds may lead to non-equilibrium
artifacts~\cite{Sarangapani2011}.)

In order to establish that our theory is general, we  analyzed
experimental results from both selectin systems
(P-selectin~\cite{marshall_direct_2003},
L-selectin~\cite{lou_flow-enhanced_2006}), and others (fibronectin
disassociating from a truncated construct of $\alpha_5 \beta_1$
integrin~\cite{kong_demonstration_2009} and myosin unbinding from
actin~\cite{guo_mechanics_2006}).  Details of the maximum-likelihood
procedure for obtaining the best-fit parameter values (Table~I) are in
the SI.  All the systems in Figs.~\ref{f3} and \ref{f4}
exhibit catch-bonds at low forces except in Fig.~\ref{f3}b, which for
comparison shows  P-selectin forming a slip-bond ($E_1=0$) with the
antibody G1.

\textit{Selectin family}:  Fig.~\ref{f3}a includes data for P-selectin with two
different forms of PSGL-1 ligand: sPSGL-1, which is a monomer
interacting with single lectin domains, and PSGL-1, which is a dimer
capable of simultaneously forming two bonds with two neighboring
lectin domains~\cite{marshall_direct_2003}.  (All other selectin
complexes, including L-selectin / PSGL-1~\cite{lou_flow-enhanced_2006}
in Fig.~\ref{f3}c-d, involve only monomeric interactions.)  We fit
both curves in Fig.~\ref{f3}a with the same set of parameters, using
$\tau(F)$ from Eq.~\eqref{e2} for the monomeric case, and in the dimer
case $\tau_\text{dim}(F) \equiv \tau(F/2) + \tau(F/2)[1+k_r
\tau(F/2)]/2$~\cite{marshall_direct_2003}.  When the dimer is intact,
each bond feels a force $F/2$. When one of the bonds break, the intact
bond still feels a force approximately equal to $F/2$, due to the
large stiffness and roughly constant displacement of the AFM
cantilever~\cite{Pereverzev2005}. In the latter case, the broken bond
can reform with some rate $k_r$, which adds one fitting
parameter. $\tau_{\text{dim}}(F)$ is a model that accounts for all
these possibilities. The resulting fits in Fig.~\ref{f3}a show that a
total of five parameters ($k_r \approx 1.1 \pm 0.3$ s$^{-1}$, the rest
listed in Table~I) can simultaneously capture the general lifetime
behaviors of both data sets.



\textit{Physical meaning of the parameters:} A \textit{sine qua non}
of a valid theory of any phenomenon is that the extracted parameters
must have sound physical meaning. In order to provide a structural
interpretation of the extracted parameters for the selectin systems,
it is instructive to compare the resulting energy and length scales to
what we know about selectin bonds independent of the model.  From the
crystal structure of the P-selectin / PSGL-1 complex in
Fig.~\ref{f1}b, we estimated that the regions B$_0$ and B$_1$ involve,
respectively, 14 and 6 ligand-lectin hydrogen bonds (the B$_1$ bonds
are shown in Fig.~\ref{f2}a). We used the software PyMol~\cite{PyMOL}
to count hydrogen bonds, with a distance cutoff of $0.35$ nm for the
heavy atoms.  The corresponding energy scales from Table~I are $E_0 =
17$ $k_BT$ and $E_1 = 9$ $k_BT$, which gives an enthalpy of $\approx
1.2-1.5$ $k_BT$ per hydrogen bond. This range is consistent with
earlier estimates of the strength of hydrogen bonds in
proteins~\cite{Bolen2008}. The distance from the EGF domain--lectin
interface to the lectin--ligand interface is $\approx 3$ nm. Since the
crystal structures suggest that restructuring of hydrogen bonds in
this region leads to catch-bond behavior, $r_0$ should be $\approx 3$
nm or less. The fitted values of $r_0 \approx 0.2 - 2.0$ nm for L- and
P-selectin, lie well within this estimate. The transition distances
$d$ vary between $\approx 0.1 - 0.6$ nm, which is the range typical
for proteins~\cite{Elms2012}.  Given the realistic values for all the
fitted parameters, our theoretical model is an accurate coarse-grained
description of selectin-type systems.

The sum $E_0+E_1$ is essentially constant for a given selectin
receptor, independent of the ligand: $E_0+E_1 \approx 27\:k_B T$ for
P-selectin and $\approx 31\:k_B T$ for L-selectin.  This suggests that
the maximum number of possible interactions is fixed by the
interactions associated with the receptor interface. For each ligand
there is a different partitioning of these interactions among those
that contribute to $E_0$ and $E_1$. The values of $E_0$ and $E_1$ can
be estimated from the structures alone using $E_0 \approx n_b
\epsilon_{hb}$ and $E_0+E_1 \approx n_e \epsilon_{hb}$ where $n_b$,
$n_e$ are the number of hydrogen bonds in the bent and extended states
respectively and $\epsilon_{hb}$ is the strength of a hydrogen bond.
For the catch-bond complexes, $E_1 \approx 7-10$ $k_B T$, or roughly
5-8 noncovalent bonds.  For P-selectin and G1 [Fig.~\ref{f3}b], all
interactions contribute to $E_0$, and we get slip-bond behavior
instead; G1 is a blocking monoclonal antibody for P-selectin.  In this
case the binding is so strong, involving all possible interactions at
the interface, that there is no room for additional stabilization
under alignment ($E_1=0$).  The finding that the ligands achieve
nearly the same value of $E_0+E_1$ means that in the aligned state
each of the considered ligands is capable of maximally exploiting the
binding partners among the receptor residues.  Our model predicts that
if the ligand were made defective, by truncating or mutating some
portion of the ligand binding sites so that their interactions with
the receptor were eliminated, the sum $E_0 + E_1$ should decrease.  We
will return to this case below in discussing a mutant of the ligand
PSGL-1.

\begin{table}[t]
\caption{Best-fit parameter values of the catch-bond model for the various experimental complexes shown in Figs.~3-4.  Parentheses denote uncertainties in the least significant digit.}
\begin{tabular}{lcccc} 
Complex & $E_1 $& $E_0 $ & $d$ & $r_0$\\[-0.25em]
& [$k_B T$] & [$k_B T$] & [nm] & [nm] \\ \hline
Psel / (s)PSGL-1 & $9.3(2)$ & $17.2(3)$ & $0.56(2)$ & $2.0(1)$ \\ 
Psel / G1 & $0$  & $26.73(4)$ & $0.51(3)$  & $2.0$ \textsuperscript{\emph{a}} \\ 
Lsel / PSGL-1& 10.2(7)&20.3(6) &0.14(4) &0.38(7) \\
LselN138G / PSGL-1 & 8.7(6)&21.8(5) & 0.14(4)&0.38(7) \\
Lsel / $\text{6-sulfo-sLe}^{\text{x}}$ & 8.7(7)&22.7(4) &0.17(4) &0.23(5) \\
LselN138G / $\text{6-sulfo-sLe}^{\text{x}}$ &7.0(7) &24.3(3) &0.17(4) &0.23(5) \\
integrin / fibronectin & $12(1)$ & $23(1)$ & $0.7(1)$ & $0.5(2)$ \\ 
actin / myosin (ADP) & $4.1(3)$ & $18.2(5)$ & $0.47(4)$ & $2.6(5)$ \\ 
actin / myosin (rigor) & $3.9(4)$ & $18.4(8)$ & $0.50(5)$ & $2.2(7)$ \\ 
\hline
\end{tabular} 

\textsuperscript{\emph{a}}{ For this Psel slip-bond system, the lack of data at small forces prevents independent fitting of $r_0$, so its value is set to the $r_0$ result for Psel/(s)PSGL-1.}
\end{table}

\textit{Integrin}: In the case of the integrin-fibronectin complex, we
took as an example AFM data for a truncated integrin (only the
headpiece of $\alpha_5 \beta_1$) binding to fibronectin FNIII$_{7-10}$
(fibronectin fragment comprising the 7-10th type III
repeats)~\cite{kong_demonstration_2009}. There is ample evidence for
an angle-dependent detachment of ligand in the integrin
headpiece~\cite{luo_structural_2007}, where the $\beta$-hybrid domain
swings out from the $\alpha$ subunit via multiple intermediate
states~\cite{zhu_complete_2013}. Our model is well suited to describe
these structural changes, and the quality of fit to experimental data
[Fig.~\ref{f4}a] shows that the physics governing the effect of force
on selectin complexes also holds for the complex involving
integrin. We can compare some of the fitted parameters with a recently
obtained crystal structure of the $\alpha_5 \beta_1$ headpiece
complexed with fibronectin (only the RGD peptide portion of
fibronectin is resolved in the structure,
Fig.~\ref{f2}b)~\cite{nagae_crystal_2012}.

Since the structure shows the integrin headpiece in a closed (large
angle) conformation, we can directly compare the number of hydrogen
bonds with the parameter $E_0$.  As shown in Fig.~\ref{f2}b, there are
nine hydrogen bonds formed between the headpiece domain and the RGD
peptide.  In addition, the acidic residue Asp forms a salt-bridge with
the ligand residue Arg.  Beyond the interactions that can be
ascertained from the crystal structure, it is also known that
additional ``synergy'' sites in the ligand, not visible in the
structure, play a role in binding.  From the measured decrease in
binding affinity of fibronectin fragments lacking the synergy sites,
their contribution to the binding energy can be estimated to be $\approx
2-4$ $k_BT$~\cite{nagae_crystal_2012}.  Combining this with the
hydrogen bonds and salt bridges seen in the structure (using our
earlier range of $1.2-1.5$ $k_BT$ per hydrogen bond, and $4-8$ $k_BT$
for the salt bridge~\cite{Gohlke2002}) we get an estimated total of
$E_0 = (17-26)$ $k_BT$.  Our fitted result $E_0 = 23$ $k_B T$ from the
model falls in this range, and is therefore consistent with the
structural analysis.  The fitted value of $r_0$ is also reasonable,
given that the longest axis of the hybrid domain is $\sim 4$ nm. The
parameter $d$ is again well within the range of transition state
distances expected in proteins.  Our model predicts that $E_1=12$ $k_B
T$, the extra interaction strength that would be gained in an open
conformation of the $\alpha_5 \beta_1$--fibronectin complex. This prediction
can be verified once crystal structures of the open
conformation become available.

\textit{Actomyosin}: Finally, in the case of actomyosin catch-bonds
[Fig.~\ref{f4}b], no crystal structures exist for the complex and an
angle-dependent lifetime has not been established. However, we can use
our theory to propose the origins of catch-bond behavior in these
complexes based on experimental data. There is strong evidence that
the upper 50K and lower 50K domains surrounding the major cleft in the
motor head behave like pincers---binding to actin tightly in the ADP
and rigor states, thereby forming a tight
complex~\cite{holmes_electron_2003}. Once ATP binds, the pincers move
apart (the upper 50K domain breaks contact with actin) by an
allosteric mechanism~\cite{Tehver10Structure}, thus allowing the motor
head to unbind from actin faster.  While in the ADP/rigor state, if an
external force is applied through the lever arms of myosin, local
rearrangements and rotations would cause the N terminal domain and the
two 50K domains to align with the direction of force. Along with these
local reorientations, the force would also stretch the domains,
causing narrowing of the major cleft, and facilitating increased
interactions of both 50K domains with actin. This mechanism would lead
to catch-bond behavior, in a manner similar to the FimH-mannose
adhesions in \textit{E. Coli}~\cite{le_trong_structural_2010}. Our
fitted value of $r_0$ shows that the alignments occur over a
length-scale $\sim 2.4$ nm, which agrees well with single molecule
results showing that the cross-bridge compliance resides only locally
in the actin-motor domain of the actomyosin
complex~\cite{molloy_single-molecule_1995}.

\noindent\textbf{Predictions for mutations in selectin complexes:} Since the
energy scales in our model correspond to the strengths of noncovalent
bonding networks, we can use our theory to predict and explain the
impact of mutations on the bond lifetime, thus providing a framework
for engineering catch-bonds with specific properties.  We will
consider two examples, one a modification of the ligand, the other of
the receptor in selectin systems.  A recent study~\cite{Xiao2012}
considered a PSGL-1 mutant where Tys51, a sulfated tyrosine that makes
one hydrogen bond with Arg85 in the B$_1$ region of P-selectin
[Fig.~\ref{f2}], is replaced by phenylalanine (Phe), which cannot form
the hydrogen bond.  Kinetic assays showed that the mutant has a weaker
binding affinity to P-selectin, but a zero force off-rate that remains
virtually unchanged from the wild-type.  The lifetime under force has
not yet been measured, but our model predicts that removing one
hydrogen bond from B$_1$ should decrease $E_1$ by $\approx 1.3$ $k_B
T$. Using the reduced value for $E_1$ with all other parameters the
same as in the wild-type (first row of Table~I), we predict that the
$\tau(F)$ curve (dashed red line labeled Y51F in Fig.~\ref{f3}a),
should be dramatically different from the wild-type.  Relative to the
wild-type, the peak is decreased by a factor of 3.4, and shifted
slightly (from 24 to 21 pN).  Since effects of a mutation in $E_1$ are
most relevant to alignment under force, the low force behavior is
relatively unperturbed, similar to the kinetic assay results:
$\tau(F)$ of the mutant for $F < 2$ pN differs less than 20\% from the
wild-type.

The second example, where experimental $\tau(F)$ data is available,
involves two receptor mutations performed on L-selectin
\cite{lou_flow-enhanced_2006}. The authors compared the
$\tau(F)$ behavior of wild-type L-selectin to a mutant where Asn138
was changed to Gly.  The mutation effectively breaks a hydrogen bond
in the hinge region, between Tyr37 and Asn138. Two different ligands
(PSGL-1 and 6-sulfo-sLe$^\text{x}$) both showed the same trends: the
peak in the $\tau(F)$ curve for the mutant was shifted up and toward
smaller forces, relative to the wild-type [Fig.~\ref{f3}c-d].  To
determine the minimal perturbation in the parameters that would
produce this shift, we simultaneously fit the wild-type and mutant
data sets for each ligand, allowing only a subset of parameters to
change for the mutant.  The most likely subset, determined using the
Akaike information criterion (see SI for details), involved only changes in the
energy scales $E_0$ and $E_1$.  The fit results are shown in Table~I.
Both ligands show a similar pattern: $E_1$ decreased by $\approx
1.5-1.7$ $k_BT$ in the mutant, while $E_0$ increased by $\approx
1.5-1.6$ $k_BT$.  The magnitudes of the energy changes suggest that
the enthalpy loss due to a single hydrogen bond contributing to $E_1$
in the wild-type is compensated by an increase in $E_0$.  The mutation
gives added flexibility to the lectin domain, allowing it to bind the
ligand more effectively in both the bent and extended conformations.
Thus, a contact between the ligand and receptor in $B_1$
[Fig.~\ref{f1}] that forms only at small angles in the wild-type, is
present at all angles in the mutant.

\section*{Conclusions}

The general principle for the formation of catch-bonds emerging from
experiments and theory is an increase in stabilizing interactions as a
result of topological rearrangements of protein domains under
force~\cite{barsegov_dynamics_2005,springer_structural_2009}. While we quantitatively establish the mechanism for certain
classes of protein complexes in this work, recent computational
studies on a knotted protein~\cite{Sulkowska09PRL} and a long $\alpha$
helix~\cite{kreuzer_catch_2013} suggest that the same principle could
lead to non-monotonic unfolding lifetimes in single proteins as
well. In the former, the protein thymidine kinase was studied, where a
``threaded'' loop is surrounded by a ``knotting'' loop, forming a
slip-knot~\cite{Sulkowska09PRL}. At intermediate forces, the knotting
loop shrinks faster than the threaded loop, effectively leading to
increased interactions between the loops and hence an increased
barrier to unfolding. At smaller forces, the threaded loop shrinks
faster and slips out of the knotting loop, before any extra
interactions can form. In the beta-myosin helix studied using molecular simulations with the milestoning algorithm~\cite{kreuzer_catch_2013}, at intermediate forces broken hydrogen bonds from the native alpha helix secondary structure reform to create a longer-lived force-stabilized pi helix structure, thereby leading to a catch-bond like effect. More generally, we suggest that if the number of hydrogen bond or side chain interactions can be increased in single domain proteins by force-induced structural rearrangements then such systems should exhibit catch bond behavior. This is likely to be the case in mammalian prions which have a number of unsatisfied hydrogen bonds in the functional state~\cite{Dima02BJ}.  Thus, it is increasingly becoming clear that diverse force-induced topological rearrangements can be used by nature as a mechanism to modulate bond lifetimes.

At the larger scale of protein complexes, one can ask whether the
rearrangements responsible for catch-bonds among different biomolecule
families share common features.  From structure-based observations in
selectin and integrin systems, we have shown that a model based on
force dependent rotation of protein domains, facilitating enhanced
interactions with their binding partner, explains experimental
observations remarkably well. Our precise analytical theory
quantitatively reproduces data on a variety of structurally unrelated
complexes with lifetimes spanning nearly four orders of magnitude.
More importantly, the key parameters of the theory are linked to the
formation (or disruption) of a network of hydrogen bonds and/or
salt-bridges. Because the strength of these interactions can be
estimated, our theory can be readily used to predict the effects of
mutations, as demonstrated for the selectin complexes. Interestingly,
analysis of experimental data allowed us to predict the strength of
additional hydrogen bonds that form in the open $\alpha_5 \beta_1$
integrin--fibronectin complex.  The specificity of our model, with
very few parameters, lays a foundation for synthetic
mechanochemistry~\cite{konda_molecular_2013}: designing and
fine-tuning catch-bond adhesion complexes with a desired set of
load-bearing characteristics.

{\bf Acknowledgements:} This work was supported by a grant from the National Institutes of Health through grant number GM 089685.


\newpage

\begin{widetext}

\renewcommand{\theequation}{S\arabic{equation}}
\renewcommand{\thetable}{S\arabic{table}}
\renewcommand{\thefigure}{S\arabic{figure}}
\setcounter{equation}{0}
\setcounter{figure}{0}
\setcounter{section}{0}

\begin{center}
{\Large Supplementary Information for:\\
 Plasticity of hydrogen bond networks regulates mechanochemistry of cell adhesion complexes}\\
Shaon Chakrabarti, Michael Hinczewski, and D. Thirumalai
\end{center}

\section{Derivation of the equation for bond lifetime}

The dynamics of our model can be described by the probability density
$\Psi(\mb{r},t)$ to find the system with bond vector $\mb{r} =
(r,\theta,\phi)$ at time $t$.  This probability evolves according to
the Fokker-Planck equation in spherical coordinates,
\begin{equation}\label{e1}
\begin{split}
\frac{\partial \Psi}{\partial t}=& \frac{D}{r^{2}}\frac{\partial}{\partial r} \left[ r^{2}e^{-\beta U }\frac{\partial{\left( e^{\beta U}\Psi   \right)}}{\partial{r}}\right]+\frac{D}{r^2\sin\theta} \frac{\partial}{\partial \theta} \left[  \sin\theta  e^{-\beta U} \frac{\partial{\left( e^{\beta U}  \Psi  \right)}}{\partial{\theta}}\right]\\
&+ \frac{D}{r^{2}\text{sin}^{2} \theta} \frac{\partial}{\partial{\phi}} \left[ e^{-\beta U} \frac{\partial{\left( e^{\beta U} \Psi \right)}}{\partial{\phi}} \right],
\end{split}
\end{equation}
with $\beta = 1/k_BT$.  Eq.~\eqref{e1} describes diffusion on the
energy surface $U(r,\theta)$,
\begin{equation}\label{e2}
U(r,\theta)=\frac{1}{2}\left( k_0+k_1(1+\cos\theta) \right)  (r-r_0)^{2}-Fr\cos\theta,
\end{equation}
with diffusion constant $D$.  We define the marginal probability
$P(r,\theta,t)$ by multiplying $\Psi$ with the spherical Jacobian and
integrating over the azimuthal angle $\phi$,
\begin{equation}\label{e3}
P(r,\theta,t) \equiv r^2 \sin\theta \int_0^{2\pi} d\phi \Psi(\mb{r},t).
\end{equation}
Since $U$ is independent of $\phi$, carrying out the operation in Eq.~\eqref{e3} and using
Eq.~\eqref{e1} leads to a two-dimensional Fokker-Planck equation for
$P(r,\theta,t)$,
\begin{equation}\label{e4}
\frac{\partial P}{\partial t}=D \frac{\partial}{\partial{r}} \left[ e^{-\beta V}\frac{\partial{\left( e^{\beta V} P   \right)}}{\partial{r}}   \right]+\frac{D}{r^2} \frac{\partial}{\partial{\theta}}
\left[  e^{-\beta V} \frac{\partial{\left( e^{\beta V}  P  \right)}}{\partial{\theta}}    \right], 
\end{equation}
in terms of a modified potential
\begin{equation}\label{e5}
V(r,\theta) = U(r,\theta) -k_BT \log
(r^2 \sin\theta).
\end{equation}
For a given force $F$, we are interested in the mean first passage time
(MFPT) $\tau_0(r,\theta,F)$ from a point $(r,\theta)$ with $r<b$ to
any point $(b,\theta^\prime)$ at the boundary defining bond rupture.
The MFPT satisfies the following
equation~\cite{kampen_stochastic_2007}, derived from the backward
Fokker-Planck equation,
\begin{equation}\label{e6}
D \frac{\partial}{\partial r}\left[e^{-\beta V} \frac{\partial \tau_0}{\partial r}\right] + \frac{D }{r^2}\frac{\partial}{\partial \theta}\left[e^{-\beta V} \frac{\partial \tau_0}{\partial \theta} \right] =-e^{-\beta V},
\end{equation}
with boundary condition $\tau_0(b,\theta^\prime,F) = 0$ for all
$\theta^\prime$.  Since the two-dimensional first-passage problem in
Eq.~\eqref{e6} cannot be solved analytically, we will approximately
map it to a one-dimensional problem.  Integrating Eq.~\eqref{e6} over
$\theta$ leads to
\begin{equation}\label{e7}
D \frac{\partial}{\partial r}\int_0^\pi d\theta\,e^{-\beta V(r,\theta)} \frac{\partial}{\partial r} \tau_0(r,\theta,F) =-\int_0^\pi d\theta\, e^{-\beta V(r,\theta)}.
\end{equation}
The second term in Eq.~\eqref{e6} vanishes under the integration because
$\exp(-\beta V(r,\theta)) \to 0$ in the limits $\theta
\to 0^+$ and $\theta \to \pi^-$, as can be seen from Eq.~\eqref{e5}.

In order to evaluate the integral on the left hand side of Eq.~\eqref{e7} we make a saddle-point approximation,  replacing $\partial \tau_0(r,\theta,F)/\partial r$ with,
$\partial \tau_0(r,\theta_\text{m}(r),F)/\partial r$, where
$\theta_\text{m}(r)$ is the location of the minimum of $V(r,\theta)$
at a fixed radius $r$.  For our potential, a single such minimum exists for
any given $r$, making $\theta_\text{m}(r)$ a well-defined function.  The
result is an approximate one-dimensional MFPT equation,
\begin{equation}\label{e8}
D \frac{\partial}{\partial r} \left[ e^{-\beta \tilde{V}(r)} \frac{\partial}{\partial r} \tilde{\tau}_0(r,F)\right]=-e^{-\beta \tilde V(r)},
\end{equation}
where $\tilde{\tau}_0(r,F) \equiv \tau_0(r,\theta_\text{m}(r),F)$ and
the effective one-dimensional potential $\tilde{V}(r)$ is given by
\begin{equation}\label{e9}
\tilde{V}(r) \equiv  -\frac{1}{\beta}\log \int_0^\pi d\theta\, e^{-\beta V(r,\theta)} = -\frac{1}{\beta}\log\left[\frac{r^2 e^{-\beta(F r + \frac{1}{2}k_0(r-r_0)^2)}\left(e^{2\beta(F r - \frac{1}{2}k_1(r-r_0)^2)}-1\right)}{\beta (F r -\frac{1}{2}k_1(r-r_0)^2)} \right].
\end{equation}
With the boundary condition $\tilde{\tau}_0(b,F) = 0$, Eq.~\eqref{e8}
can be solved for $\tilde{\tau}_0(r,F)$,
\begin{equation}\label{e10}
\tilde{\tau}_0(r,F) = \frac{1}{D} \int_r^b dr^\prime\,e^{\beta \tilde V(r^\prime)} \int_0^{r^\prime} dr^{\prime\prime} e^{- \beta \tilde V(r^{\prime\prime})}.
\end{equation}
The function $\tilde{V}(r^\prime)$ is a monotonically increasing
function of $r^\prime$ at large $r^\prime$.  Hence the integral over
$r^\prime$ in Eq.~\eqref{e10} gets its dominant contribution from
$r^\prime$ near the upper limit $b$, due to the presence of the
$\exp(\beta \tilde V(r^\prime))$ term.  To simplify the integral, we
will make two approximations: (i) Expand $\tilde V(r^\prime)
\approx \tilde V(b) + \tilde{V}^\prime(b) (r^\prime - b)$.  (ii)
Assume $b \gg r_\text{m}$, where $r_\text{m}$ is the location of the
minimum in $\tilde V(r)$, so that the upper limit in the inner integral over
$r^{\prime\prime}$ can be replaced by $\infty$.  If the initial
position $r$ is close to the potential minimum at $r_\text{m}$, so that $b
\gg r$, the integrals in Eq.~\eqref{e10} can be then approximately
carried out to yield
\begin{equation}\label{e11}
\tilde{\tau}_0(r,F) \approx \frac{e^{\beta \tilde{V}(b)}}{\beta D \tilde{V}^\prime(b)} \int_0^\infty dr^{\prime\prime}\,e^{-\beta \tilde{V}(r^{\prime\prime})} = \left[D \tilde{P}^\prime(b) \right]^{-1},
\end{equation}
where
\begin{equation}\label{e12}
\tilde{P}(r) \equiv {\tilde Z}^{-1} e^{-\beta \tilde V(r)}, \quad \tilde{Z} \equiv \int_0^\infty dr^\prime\,e^{-\beta \tilde V(r^\prime)}.
\end{equation}
Since under this approximation $\tilde{\tau}_0(r,F)$ is independent of
the starting point $r$, we will drop the $r$ dependence, and simplify
the notation by defining the approximate bond lifetime $\tau(F) \equiv
\tilde{\tau}_0(r,F)$.

To obtain an analytical expression for $\tau(F)$, we need to evaluate
the integral for $\tilde{Z}$ in Eq.~\eqref{e12} for $\tilde{P}(r)$.
Since this cannot be done exactly, we will approximate $\tilde{Z}$ as a
Gaussian integral by expanding $\tilde{V}(r)$ around $r=r_\text{m}$ to
second order, leading to
\begin{equation}\label{e13}
\tilde{Z} \approx \left(\frac{\beta\tilde{V}^{\prime\prime}(r_\text{m})}{2\pi}\right)^{-1} e^{-\beta \tilde{V}(r_\text{m})}.
\end{equation}
To find closed-form expressions for $r_\text{m}$ and the
$\tilde{V}^{\prime\prime}(r_\text{m})$, we note that the location of
the minimum of $\tilde{V}(r)$ and the curvature at the minimum
approximately coincide with those of the simpler potential
$\tilde{V}_\text{s}(r)$,
\begin{equation}\label{e14}
  \tilde{V}_\text{s}(r) = \frac{1}{2}\left( k_{0}+2k_{1} \right)  (r-r_0)^{2}-F r -2 k_b T \log r,
\end{equation}
which comes from substituting $\cos(\theta) \to 1$ in $V(r,\theta)$ in
the integral defining $\tilde{V}(r)$ [Eq.~\eqref{e9}].  Fig.~\ref{S1}
illustrates $\tilde{V}(r)$ versus $\tilde{V}_\text{s}(r)$ at two

\begin{figure}[t]
\includegraphics[width=0.85\textwidth]{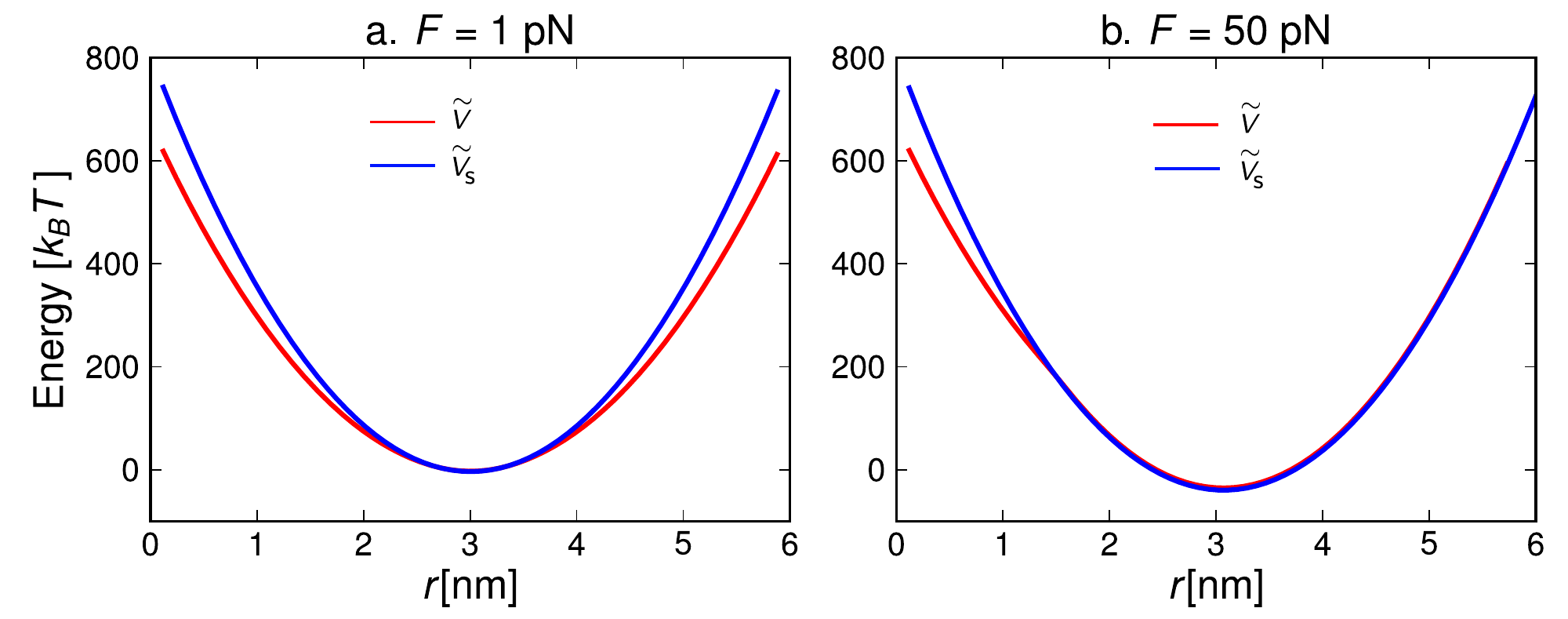}
\caption{Comparison of the potentials $\tilde{V}(r)$ and
  $\tilde{V}_\text{s}(r)$ at two different forces: a) $F = 1$ pN; b)
  $F=50$ pN.  The energy scales are aligned such that the minima of
  both potentials occur at $0$ $k_BT$.  The parameters are:
  $k_0=147.2\: k_BT/\text{nm}^2$, $k_1=15.6\: k_BT/\text{nm}^2$,
  $r_0=3.0$ nm.}\label{S1}
\end{figure}

different $F$. Obtaining the location and curvature of the minimum using the simple potential $\tilde{V}_\text{s}(r)$ is justified because of the following observations: The exact location of the minimum $r_\text{m}$, is always very close to $r_0$. At zero external force or forces very close to zero, $V(r,\theta)$ is approximately the same as the simpler potential obtained by setting $\cos(\theta) \to 1$ in $V(r,\theta)$, in regions $r \sim r_0$. Hence, $\tilde{V}(r)$ and $\tilde{V}_{\text{s}}(r)$ will be similar around $r=r_0$. At larger forces, $V(r,\theta)$ and its simpler version are approximately the same only around $r \sim r_0$ and $\theta \sim 0$.  However, since $V(r,\theta)$ is minimized around $\theta \sim 0$ in regions around $r_0$, the dominant contribution to the integral in Eq.~\eqref{e9} for $r$ values around $r_0$ comes from $\theta \sim 0$. Hence once again the simpler form of $V(r,\theta)$ can be used leading to similar $\tilde{V}(r)$ and $\tilde{V}_{\text{s}}(r)$ around $r=r_0$. The potential $\tilde{V}_\text{s}(r)$ reaches its
minimum at
\begin{equation}\label{e15}
r_\text{ms} = 4 \left[-\beta(F+(k_0+2k_1)r_0) + \sqrt{8\beta(k_0+2k_1)+\beta^2(F+(k_0+2k_1)r_0)^2} \right]^{-1},
\end{equation}
where the curvature is given by
\begin{equation}\label{e16}
\tilde{V}^{\prime\prime}_\text{s}(r_\text{ms}) = k_0+2k_1 +\frac{2}{\beta r_\text{ms}^2}.
\end{equation}
The complete approximation for $\tilde{Z}$ involves substituting
Eqs.~\eqref{e15} and \eqref{e16} for $r_\text{m}$ and
$\tilde{V}^{\prime\prime}(r_\text{m})$ in Eq.~\eqref{e13},
\begin{equation}\label{e17}
\tilde{Z} \approx \left(\frac{\beta\tilde{V}^{\prime\prime}_\text{s}(r_\text{ms})}{2\pi}\right)^{-1} e^{-\beta \tilde{V}(r_\text{ms})}.
\end{equation}

Plugging the definition of $\tilde{V}(r)$ from Eq.~\eqref{e9} and
$\tilde{Z}$ from Eq.~\eqref{e17} into Eq.~\eqref{e12} for
$\tilde{P}(r)$, we can now analytically approximate $\tau(F) = [D
\tilde{P}^\prime(b) ]^{-1}$.  The resulting expression simplifies for
large $k_0$, corresponding to large energy barriers for bond rupture,
yielding the final form for the bond lifetime [Eq.~(2) of the main
text],
\begin{equation}\label{e18}
\tau (F)\approx \frac{ \sqrt{\pi}
    \, r_0 (E_1 - 2
    F (d + r_0)) e^{\beta(E_0+d F)} (e^{2 \beta F r_0}-1)}{ 4 D (\beta E_0)^{3/2} F \left(1+ r_0/d \right) ^2 (1 - e^{\beta (2F (d + r_0)-E_1)})},
\end{equation}
where $E_0 = k_0 d^2/2$ and $E_1= k_1 d^2$.

\section{Brownian Dynamics Simulations}

To check the accuracy of the theoretical prediction for the lifetime $\tau(F)$ in
Eq.~\eqref{e18}, we performed overdamped Brownian dynamics
simulations~\cite{Ermak1978} for a test particle of radius $r_0$ diffusing in
the potential $U$ given in Eq.~\eqref{e2} using $D =
k_BT/(6 \pi \eta r_0)$, where $\eta = 0.89$ mPa$\cdot$s is the
viscosity of water at $T=298$ K.  We chose the time step for numerical
integration to be about $2 \times 10^{-6} r_0^2/D$. The trajectories
were started with the bead at $\mb{r}_\text{min}$, the minimum of the
potential $U$, and stopped when the bead reached the rupture boundary
at $r=b$ for the first time. Statistics were obtained from $\approx 150-300$ trajectories,
depending on the value of force, and error bars on the simulated data were
estimated by the jackknife method~\cite{Miller1974}.  Fig.~\ref{S2}
shows a comparison of the numerical results to the analytical formula
of Eq.~\eqref{e18} for parameters corresponding to the rigor
actomyosin experimental system (main text Table I).  The
excellent agreement validates the approximations used to derive
Eq.~\eqref{e18}.

\begin{figure}[t]
\includegraphics[width=0.65\textwidth]{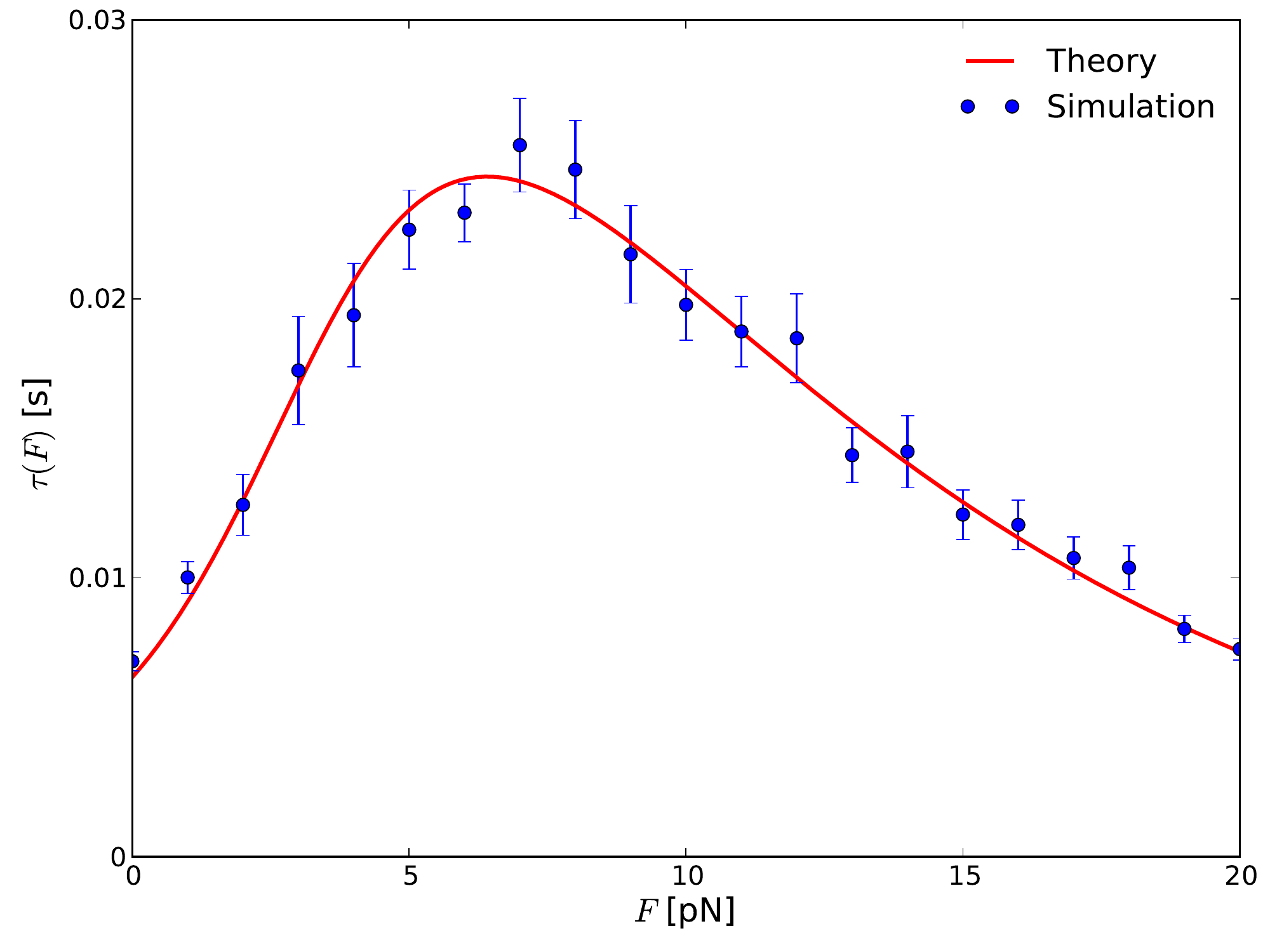}
\caption{Approximate theoretical bond lifetime $\tau(F)$
  [Eq.~\eqref{e18}, solid curve] versus the numerical results of
  Brownian dynamics simulation (circles), for parameters: $E_0 = 18.4$
  $k_BT$, $E_1 = 3.9$ $k_BT$, $d = 0.5$ nm, $r_0 = 2.2$ nm.}\label{S2}
\end{figure}

\section{Fitting to Experimental Data}

We fitted Eq.~\eqref{e18} for $\tau(F)$ to experimental data by the
standard method of minimizing $\chi^2$ values, which is equivalent to
maximizing a log-likelihood function, with the assumption that errors in the mean lifetime data are 
Gaussian-distributed.  For the fits
in Fig.~3c-d and Fig.~4 of the main text, the standard deviation for
each lifetime was obtained from the error bars given in the
corresponding experimental studies.  However, since error bars were not
provided for the lifetime data in Fig.~3a-b, we derived error bars
from the scatter in the three reported estimates for $\tau(F)$:
average lifetimes, standard deviation of the lifetimes, and -1/slope
in the logarithmic plot of the number of events with lifetime $t$ or
greater versus $t$.  For exponentially distributed lifetimes (the case in
all the experimental systems under consideration), these three
quantities should be equal to $\tau(F)$ up to deviations due to
sampling errors.  After fitting, the uncertainties in the parameters
$E_0$, $E_1$, $d$, and $r_0$ listed in Table I of the main text were
obtained from the diagonal elements of the best-fit covariance matrix.


For the simultaneous fitting of L-selectin mutation
data~\cite{lou_flow-enhanced_2006} in Fig.~3c-d of the main text, we used
the following procedure to determine the minimal perturbation to the
parameters of the system that produces the observed shift in the
$\tau(F)$ curves.  The data alone suggests that
not all the model parameters are relevant to the mutation.  The
experimental $\tau(F)$ curves for the wild-type (WT) and the mutant in Fig.~3c-d
show that the decay in $\tau(F)$ at large $F$ is similar.  Since the decay is controlled
by the parameter $d$, we assume that the  value of $d$ for the WT and the mutant
is the same.  This leaves three parameters, $E_0$, $E_1$, $r_0$, that could
potentially be altered by the mutation, though it is possible that
only a subset of these is sufficient to explain the shift. We carried
out simultaneous fitting of the model to the WT and mutant 
$\tau(F)$ curves for each ligand, under eight different hypotheses,
corresponding to different subsets of the three parameters
varying under mutation.  For a given ligand, the mutant and WT
share all parameters except the subset that is allowed to vary (first 
column of Table~S1). Between curves for different ligands, all 
parameters are distinct.  The table shows
the resulting $\chi^2$ statistic (the total $\chi^2$ for the data sets
involving both ligands). The lowest $\chi^2$ is achieved for
hypothesis 3, where all three parameters are allowed to vary.
However, this could be the result of overfitting, since hypothesis 3
also has the largest number of free parameters.  A better way to rank
the hypotheses is through the corrected Akaike information criterion,
\begin{equation}\label{e19}
\text{AICc} = \chi^2 + 2p + \frac{2p(p+1)}{n-p-1},
\end{equation}
where $n$ is the number of data points and $p$ the number of free
parameters~\cite{Burnham2002}.  The AICc penalizes overfitting due to
an excessive number of parameters, and has a natural probabilistic
interpretation: if two model fits have AICc values of $a_1$ and $a_2$
respectively, with $a_1 < a_2$, then model 2 has a likelihood
$\exp((a_1-a_2)/2)$ of being the true interpretation of the data,
relative to model 1.  From AICc values listed in Table~S1, we
see that the most likely hypothesis is 1, where $E_0$ and $E_1$ are
allowed to vary.  Hypothesis 2 ($E_1$ and $r_0$ varying) is a close
competitor (78\% as likely as 1), and the remaining ones are
increasingly improbable (hypothesis 3 is only 3\% as likely as 1).  As
argued in the main text, hypothesis 1 also has a very reasonable
physical interpretation, with the mutation causing a single bond to
switch between the sets that contribute to $E_1$ and $E_0$.
Hypothesis 2, which involves the mutation decreasing $E_1$ and
increasing the lever arm distance $r_0$, is more difficult to explain in
physical terms, but cannot be completely ruled out based on fitting
alone.  The fit results for hypothesis 1 are shown in Fig.~3c-d, and the
parameters are listed in Table I of the main text.

\begin{table}[t]
  \caption{Simultaneous fitting of the L-selectin mutation data~[4]. The first column lists eight hypotheses, corresponding to different subsets of parameters that are allowed to vary between the fits to the wild-type and mutant data sets.  $\chi^2$ is a measure of goodness of fit, and AICc is the corrected Akaike criterion.    The hypotheses are
    ordered by increasing AICc.The lowest values of $\chi^2$ and AICc are in bold. }\label{t1} 
\begin{tabular}{lccc} 
 Varying subset& $\chi ^2$ &AICc  \\ \hline
 1: $E_0$, $E_1$ & 32.2& {\bf 71.8}  \\ 
 2: $E_1$, $r_0$ & 32.7& 72.3  \\ 
 3: $E_0$, $E_1$, $r_0$ & {\bf 27.3}& 78.7   \\
 4: $E_0$, $r_0$ & 50.9& 90.5 \\  
5: $r_0$ & 65.9 & 95.9 \\ 
6: $E_0$& 90.8 & 120.8 \\ 
7: $E_1$ & 154.0 & 184.0  \\ 
8: none & 224.1 & 246.0 \\
\end{tabular} 
\end{table}


\end{widetext}

\end{document}